\documentclass[amsmath,amssymb,prl,hyperlink,twocolumn]{revtex4}

\usepackage{graphicx}
\usepackage{soul}
\usepackage[colorlinks=true,citecolor=blue,linkcolor=magenta]{hyperref}
\usepackage[usenames]{color}
\usepackage{amsfonts}

\def \be{\begin{equation}}
\def \ee{\end{equation}}
\def \ba{\begin{array}}
\def \ea{\end{array}}
\def \bea{\begin{eqnarray}}
\def \eea{\end{eqnarray}}

\newcommand{\ssm}{\scriptscriptstyle\rm}
\renewcommand{\phi}{\varphi}
\begin{document}

\title{Observation of eight-photon entanglement}

\author{\textsc{Xing-Can Yao$^{1}$, Tian-Xiong Wang$^{1}$, Ping Xu$^{1}$, He Lu$^{1}$, Ge-Sheng Pan$^{1}$, Xiao-Hui Bao$^{1,2}$, Cheng-Zhi Peng$^{1}$, Chao-Yang Lu$^{1}$, Yu-Ao Chen$^{1}$, Jian-Wei Pan$^{1,2}$}\vspace{0.2cm}}

\affiliation{$^1$ Hefei National Laboratory for Physical Sciences at Microscale and Department of Modern Physics, University of Science and Technology of China, Hefei, Anhui 230026, China}
\affiliation{$^2$ Physikalisches Institut, Ruprecht-Karls-Universit\"{a}t Heidelberg, Philosophenweg 12, 69120 Heidelberg, Germany}

\date{\vspace{0.1cm}\today}

\iffalse
\maketitle

\begin{affiliations}
 \item Hefei National Laboratory for Physical Sciences at Microscale and
Department of Modern Physics, University of Science and Technology
of China, Hefei, Anhui 230026, China
 \item Physikalisches Institut, Ruprecht-Karls-Universit\"{a}t
Heidelberg, Philosophenweg 12, 69120 Heidelberg, Germany
\end{affiliations}

\fi

\begin{abstract}
Using ultra-bright sources of pure-state entangled photons from parametric down conversion, an eight-photon interferometer and post-selection detection, we demonstrate the ability to experimentally manipulate eight individual photons and report the creation of an eight-photon Schr\"odinger cat state with an observed fidelity of $0.708 \pm 0.016$.
\end{abstract}

\pacs{}

\maketitle

The creation of increasingly large multipartite entangled states is not only a fundamental scientific endeavor itself \cite{Schrodinger35,EPR35,Leggett08}, but also the enabling technology for quantum information \cite{Zoller05,Ladd10}. Since the first experimental demonstration of entanglement among three spatially separated photons \cite{Bouwmeester99} in 1999, tremendous effort has been devoted to generating multiparticle entanglement with a growing number of qubits \cite{Sackett00,Zhao04,Haffner05,Lu07,Prevedel09,Wieczorek09,Radmark09,Matthews10,Krischek10,Monz11}. So far, the maximal number of entangled photons has been limited to six \cite{Lu07,Prevedel09,Wieczorek09,Radmark09,Matthews10} based on spontaneous parametric down-conversion \cite{Kwiat95}. Here, using new ultra-bright sources of entangled photon pairs \cite{Kim03}, an eight-photon interferometer and post-selection detection, we demonstrate for the first time the ability to experimentally manipulate eight individual photons. We create and characterize an eight-photon Schr\"odinger-cat state \cite{Schrodinger35} with genuine multipartite entanglement verified. Our eight-photon source opens the way to experimental investigations of, for example, photonic quantum simulation \cite{Lanyon10,Ma11}, topological error correction \cite{Raussendorf07}, and test of robustness of different entangled states under decoherence \cite{Barreiro10}.\vspace{2mm}

In our experiment, we aim to create eight-photon Schr\"odinger cat states \cite{Schrodinger35}, which are also known as Greenberger-Horne-Zeilinger states (GHZ) \cite{GHZ90} and can be written in the form:
\be
\left|
{SC_8}\right\rangle  = \frac{1}{{\sqrt 2 }}(\left| H
\right\rangle ^{ \otimes 8} + \left| V \right\rangle ^{ \otimes 8}
).
\label{eq:GHZ}
\ee
where $H$ and $V$ denote the horizontal and vertical polarization of the single photons. It involves an equal superposition of the eight individual photons being in two opposite polarization states. To this end, we first prepare four pairs of polarization-entangled photons in the state $\left |\phi^{+}\right\rangle=\left(\left|H\right\rangle\left|H\right\rangle+\left|V\right\rangle\left|V\right\rangle\right)/\sqrt{2}$  using parametric down-conversion (PDC) \cite{Kwiat95}, and use an eight-photon interferometer to combine them into the Schr\"odinger-cat state (\ref{eq:GHZ}).

\begin{figure*}[t!]
\includegraphics[width=0.78\linewidth]{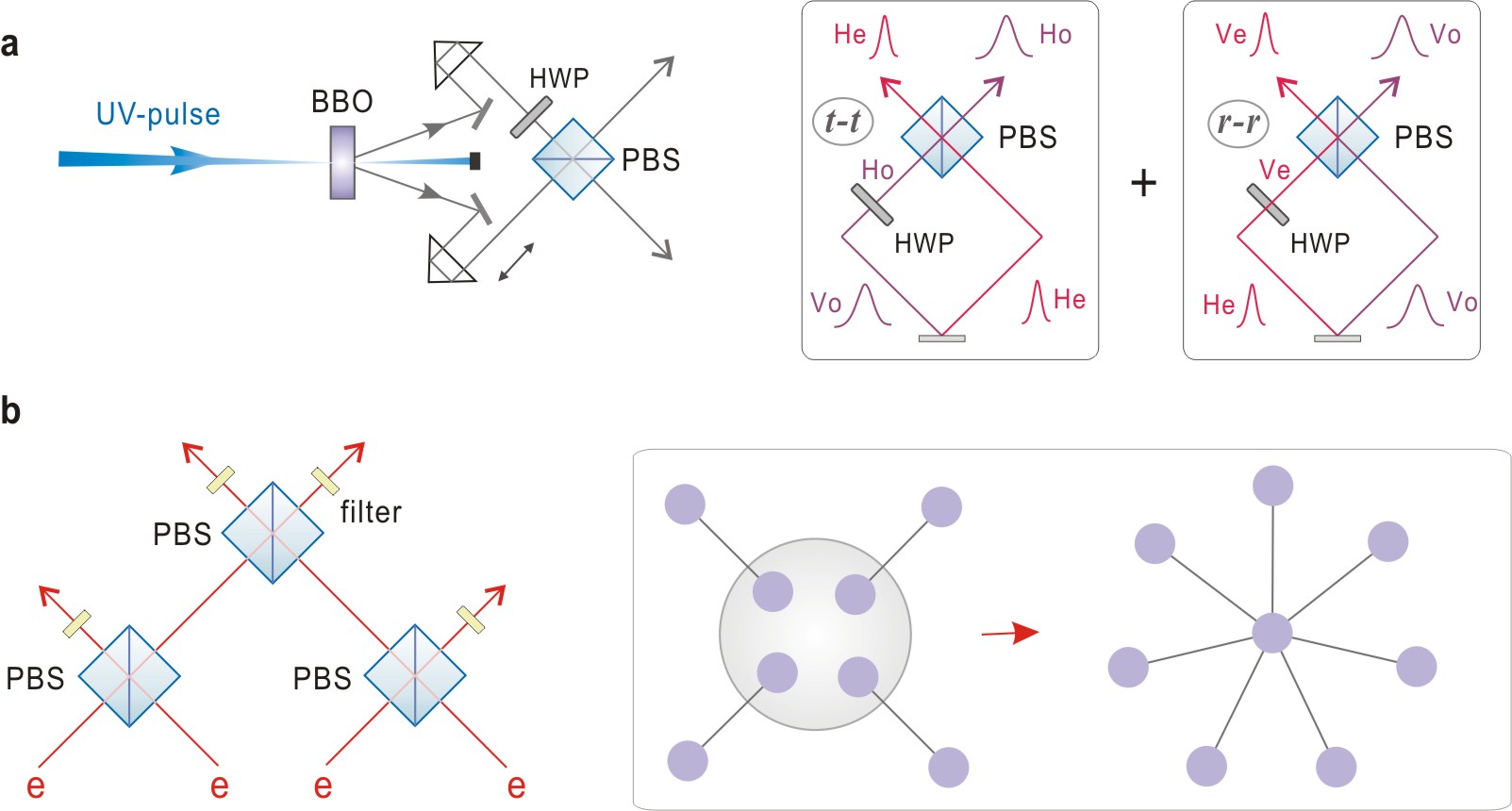}
\vspace{0cm}\caption{Experimental scheme for generating eight-photon Schr\"odinger-cat states. \textbf{a} Left panel: An initial photon pair is generated by noncollinear type-II PDC \cite{Kwiat95} and passes through a pair of birefringent compensators (not shown) which consists of a 1-mm BBO crystal and an HWP. After one of the photons's polarization is rotated by 90 degrees using an HWP, the two photons are superposed on a PBS. Right panel: the principle of an interferometric Bell-state synthesizer \cite{Kim03}. Here, the photon that leaves the BBO crystal with $e$ and $o$ polarizations with different spectral width are separated by the PBS and detected by different detectors. This effectively disentangles the timing information from the polarization information of the photon pair; generating high-fidelity entangled photons. \textbf{b} Left panel: An interferometer combining four incoming $e$-polarized photons (each from an entangled pair) with three PBSs. Upon a coincidence detection, the four pairs of entangled photons are transformed into the eight-photon Schr\"odinger-cat state (\ref{eq:GHZ}). Right panel: Graph state representation of the process of engineering the four photon pairs into the eight-photon cat state. The graph state can be thought of as being constructed by first preparing the qubits at each vertex in the state $\left|+\right\rangle=\left(\left| H\right\rangle+\left |V\right\rangle\right)/\sqrt{2}$  and then applying controlled phase gates between pairs of neighbouring qubits.  \label{Fig1}}
\end{figure*}

The first challenge of this experiment is the very low eight-photon coincidence count rate: detecting eight photons requires all the four independent pairs of entangled photons to be present at the same time, thus the eight-photon coincidence event scales as $(p\times\xi)^4$  with $p$  being the down-conversion probability and $\xi$  being the overall collection and detection efficiency, which drops quickly for small  $p$  and $\xi$. With the  $(p\times\xi)$  data from the previous six-photon experiments \cite{Lu07}, one can only expect an eight-photon event count rate of $\sim2.8\times10^{-5}$  Hz which is experimentally unfeasible. This demands a considerable improvement on the brightness of entangled photons.

The second experimental challenge is the noise control in the generation of the multi-photon entangled state. While increasing the single-pair generation rate $p$  (thus the two-photon count rate) can be straightforward by using a higher-power pumping laser, it will inevitably result in a higher double pair emission rate $\sim p^2$, which has been considered to be the main source of noise in multi-photon experiments \cite{Krischek10,Barbieri09,Weinhold08}. We will therefore need to keep the pumping laser power at a moderate regime. Later we shall also discuss the eight-photon interferometer and post-selection detection arrangement that are designed to mitigate the high-order emission noise.

To obtain entangled photon sources with both high count rate and high fidelity, here we adopt the scheme of Bell-state synthesizer proposed by Kim et al. (ref. \cite{Kim03}). As shown in Fig.\ref{Fig1}a, a type-II  $\beta$-barium borate (BBO) crystal is pumped by a femtosecond laser aiming at generating entangled photon pairs  $\left(\left|H\right\rangle\left|V\right\rangle+\left|V\right\rangle\left|H\right\rangle\right)/\sqrt{2}$. However, in the ultrafast type-II PDC, there are two types of undesired timing information correlated to the polarization which degrades the purity of entanglement. The first one is that the group velocities experienced by the different polarizations are not the same; this can be eliminated using a pair of birefringent compensators  \cite{Kwiat95}. The second one, which only occurs in ultrafast pulsed PDC, is that the $H$ and $V$ polarized light are different in their spectral (and temporal) width \cite{Keller97,Grice97}. Most previous multi-photon experiments  \cite{Lu07,Prevedel09,Wieczorek09,Radmark09,Matthews10} have relied on passing the PDC photons through narrow-band filters to select only the part of the most entangled photons (except for ref. \cite{mosley08} where heralded pure-state single photons were generated by controlling the modal structure of collinear PDC photon pair emission). Such a passive filtering process, however, is inefficient as many photons are unnecessarily wasted.

This problem can be circumvented using the interferometric Bell-state synthesizer (see Fig.\ref{Fig1}a). The photon pairs are first guided through two birefringent compensators to remove the walk-off effects, and then superposed on a polarizing beam splitter (PBS) with their path length finely adjusted to achieve perfect temporal overlapping. A half-wave plate (HWP) inserted in one arm rotates the polarization by 90$^0$ and ensures that the photon pairs have the same polarization when they reach the PBS. As the PBS transmits $H$ and reflect $V$ polarizations, there are two possible outcomes: both photons are transmitted (t-t) or both are reflected (r-r), as displayed in the right panel of Fig.\ref{Fig1}a. The PDC photons originally with $e$ polarization (with smaller spectral bandwidth) and $o$ polarization (with larger spectral bandwidth) are now separated at the exit ports of the PBS and detected by different detectors. Therefore, the timing information in the ultrafast type-II PDC cannot be used to distinguish between the t-t and the r-r paths; subsequently, the two polarization state amplitudes,  $\left|H\right\rangle\left|H\right\rangle$ and  $\left|V\right\rangle\left|V\right\rangle$, become quantum mechanically indistinguishable and form a coherent superposition state  $\left |\phi^{+}\right\rangle=\left(\left|H\right\rangle\left|H\right\rangle+\left|V\right\rangle\left|V\right\rangle\right)/\sqrt{2}$ . This effectively disentangles the timing information from the polarization information of the photon pair. Also note that since it is never the case that two photons exit the same port of the PBS, in principle no post-selection is required.

\begin{figure*}[t!]
\includegraphics[width=0.95\linewidth]{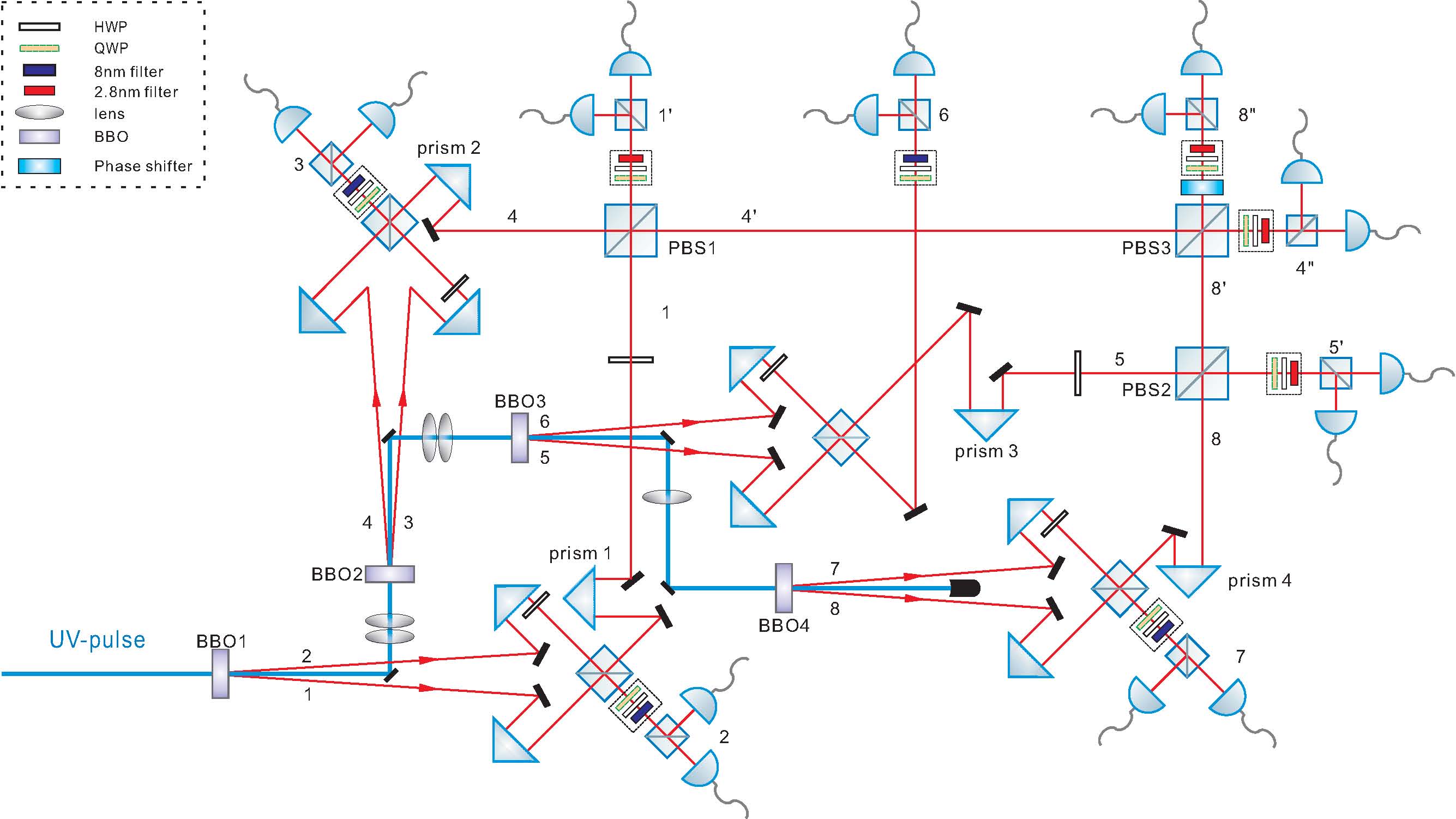}
\vspace{0cm}\caption{Experimental setup. Ultravoilet laser pulses with central wavelength of 390 nm, pulse duration of 120 fs and repetition rate of 76 MHz successively pass through four BBO crystals to produce four PDC photon pairs, which are further combined on four interferometric Bell-state synthesizers each of which consists of an HWP and a PBS. The distance between the first and the last BBO crystal is $\sim$ 1.3 m. The photon in mode 1 and 4 are then combined on the PBS1, photon 5 and 8 on PBS2, and finally photon 4$^{\prime}$ and 8$^{\prime}$ on PBS3. Technically, much effort is made to ensure good spatial and temporal overlap in the seven interferences involved in this setup, and keep them stable in a temperature stabilized laboratory. We use high-precision PBS with extinction ratio of $>$1000:1 and beam deviation of $<$ 3$^{\prime\prime}$ . Different lens settings are used for beam profile matching when overlapping on the PBSs. The photons are detected by 16 single-photon detectors (with quantum efficiency $>$ 60\%), and a complete set of the 256 eight-fold coincidence events are simultaneously registered by a homemade FPGA-based coincidence unit.  \label{Fig2}}
\end{figure*}

Next, we engineer these photon pairs into the eight-photon Schr\"odinger-cat state (\ref{eq:GHZ}). Four $e$-ray photons, each from an entangled pair, are combined on a linear optical network consisting of three PBSs, as shown in Fig.\ref{Fig1}b. It can be checked that only if all the four incoming photons have the same polarization can they be either transmitted ($\left|HHHH\right\rangle$) or reflected ($\left|VVVV\right\rangle$)  by the three PBSs, leading to a coincidence registration of a single photon at each output. Furthermore, there is no way, even in principle, to distinguish between these two possibilities,  $\left|HHHH\right\rangle$ and $\left|VVVV\right\rangle$, if all the other information (\textit{e.g.}, time, frequency, spatial mode) are erased. Thus this interferometer (Fig.\ref{Fig1}b) effectively projects (with an operator: $\left|HHHH\right\rangle\left\langle HHHH\right|+\left|VVVV\right\rangle\left\langle VVVV\right|$) the four entangled-photon pairs into the eight-photon Schr\"odinger-cat state (\ref{eq:GHZ}).

There are some interesting features of this interferometer. In graph-state picture \cite{Hein04}, it can be intuitively represented as a process of photonic qubit fusion with a star topology (see the right panel of Fig.\ref{Fig1}b). Compared to the chain topology used previously in ref. \cite{Lu07}, it allows interference only among the $e$-ray PDC photons, eliminating the bandwidth mismatch problem between the $o$- and $e$-ray photons. Technically, it also reduces noise contribution from the higher-order PDC photon emission events (see Appendix). Finally, we note that it can also be used as a four-photon GHZ state projector that allows future experimental investigations of multi-particle entanglement swapping.

Figure \ref{Fig2} shows the setup for our experiment. We use a pulsed ultraviolet laser with a power of 880 mV to successively pump four BBO crystals. Different lens settings are used to ensure that the laser is focused on each BBO with the same waist. With the use of the Bell-state synthesizer (Fig. \ref{Fig1}a), we observed an average two photon coincidence count rate of $\sim$1 MHz with fidelity of $\sim$ 90\% without the use of any narrow-band filters. Using a spectrometer, we have measured the spectral linewidth of the PDC photons; the full width at half maximum ($\lambda_{\ssm FWHM}$) is about 6nm and 12nm for the $e$- and $o$-ray photons respectively.

\begin{figure*}[t!]
\includegraphics[width=\linewidth]{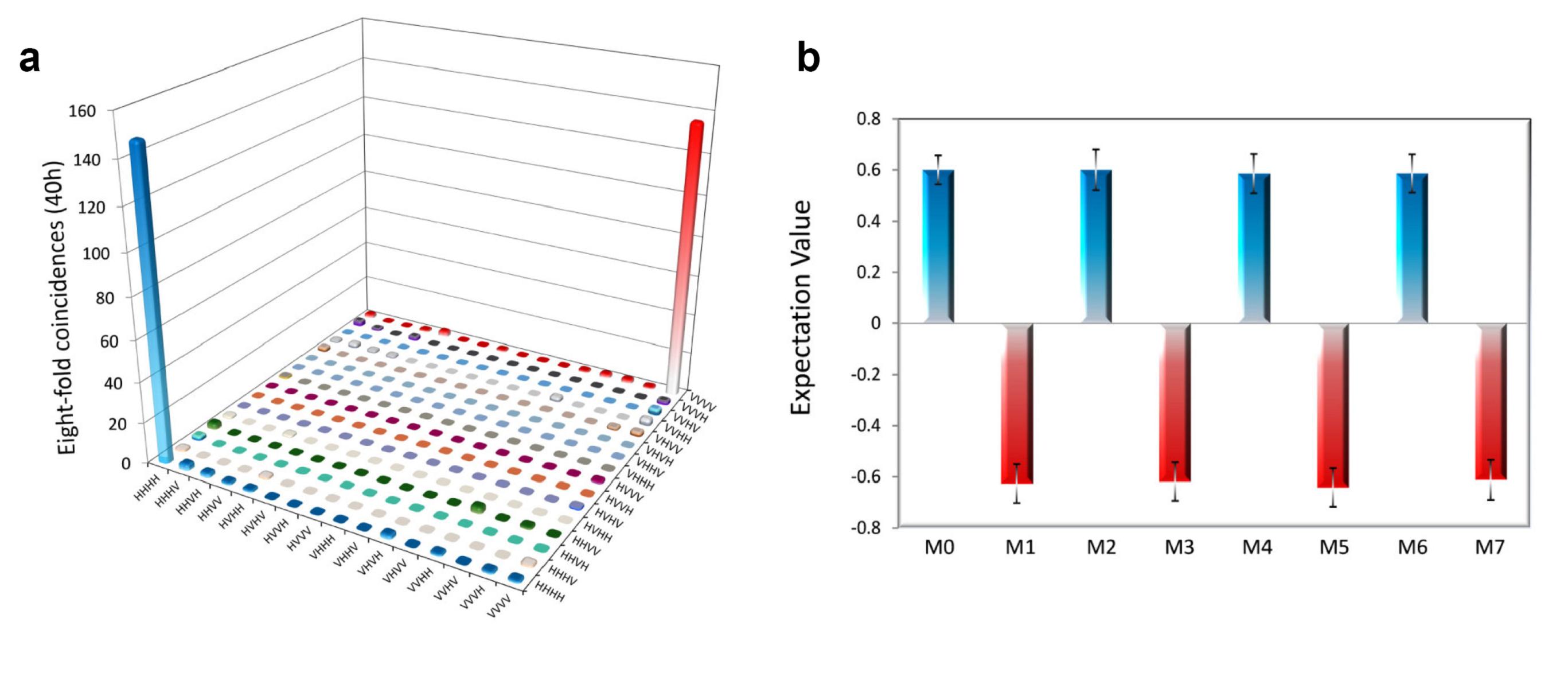}
\vspace{0cm}\caption{Experimental results for the eight-photon Schr\"odinger-cat state. \textbf{a} Coincidence counts measured in the $\left|H\right\rangle/\left|V\right\rangle$ basis accumulated for 40 hours. \textbf{b} The expectation values of $M_k^{\otimes8}$ , each derived from a complete set of 256 sixfold coincidence events in the basis of $\left|H\right\rangle\pm e^{ik\pi/8}\left|V\right\rangle$. The setting of $M_0$  is measured in 25 hours, and the rest seven settings are measured in 15 hours. The error bar stands for one standard deviation deduced from propagated Poissonian counting statistics of the raw detection events. \label{Fig3}}
\end{figure*}

Having obtained the high-brightness entangled photon source, now we overlap the $e$-ray photons 1-4-5-8 on three PBSs with good spatial mode matching, as illustrated in Fig.\ref{Fig2}. Fine adjustments of the delays between the different paths are made to ensure that the photons arrive at the PBSs simultaneously. Further, the $e$-ray photons are spectrally filtered by narrow-band filters (peak transmission rate $>$ 98\%) with $\lambda_{\ssm FWHM}=2.8$ nm, and the $o$-ray photons (2-3-4-7) are filtered by filters with $\lambda_{\ssm FWHM}=8$ nm \cite{Bouwmeester97}. With this filter setting, we observe an average two-photon coincidence count rate of $\sim$310,000 per second with a visibility of $\sim$ 94\% in the $\left|\pm\right\rangle=\left(\left|H\right\rangle\pm\left|V\right\rangle\right)/\sqrt 2$ basis. We estimate a single pair generator rate of $p=0.058$, and a collection and detection efficiency of $\xi=0.265$. To test the indistinguishability between the independent photons, we observe Hong-Ou-Mandel type interferences for the photons overlapping on the three PBSs with an average visibility of $\sim$ 76\%.

To analyze each photon's polarization state, we use a combination of a quarter-wave plate (QWP), an HWP, and a PBS, accompanied by two single-mode fibre coupled single-photon avalanche photodiodes. The obtained signals of the 16 detectors are fed into a homemade FPGA-based programmable coincidence logic unit. This unit registers a complete set of the 256 possible combinations of eight-photon coincidence events (if and only if the eight detectors in each output mode fire simultaneously), which are used for the analysis of the eight-photon cat state. However, the detected nine-photon coincidence events are discarded, thus reducing noise from higher-order PDC emissions.

In our experiment, we are able to obtain $\sim$ 9 eight-photon coincidence events per hour. To verify that the eight-photon Schr\"odinger cat stat (\ref{eq:GHZ}) has been created, we first show that under the condition of registering eight-photon coincidences, only the $\left| H \right \rangle ^{ \otimes 8}$ and $\left| V \right\rangle ^{ \otimes 8}$  are observed, but no others. This was done by comparing the counts of all 256 possible polarization combinations in the $H/V$ basis. The experiment data are shown in Fig. \ref{Fig3}a where the $\left| H \right \rangle ^{ \otimes 8}$  and $\left| V \right\rangle ^{ \otimes 8}$  terms dominate the overall coincidence events with a signal-to-noise ratio (defined as the ratio of the average of the desired components to that of the other non-desired ones) of 530:1 . Second, to verify that the $\left| H \right\rangle ^{ \otimes 8}$   and $\left| V \right\rangle ^{ \otimes 8}$   are indeed in a coherent superposition, we further perform measurements in the basis of $\left(\left|H\right\rangle\pm e^{i\theta}\left|V\right\rangle\right)/\sqrt 2$, where $\theta=k\pi/8$, $k=0,1,...,7$. It is easy to check for example, for  $k=0$, the cat state (\ref{eq:GHZ}) can be rewritten in an expression containing 128 (out of 256) terms, where those with an even number of the $\left|-\right\rangle$  component (e.g.  $\left| + \right\rangle ^{ \otimes 8}$ ,  $\left| + \right\rangle ^{ \otimes 6}\left| - \right\rangle ^{ \otimes 2}$) occur while the combinations with odd number of $\left|-\right\rangle$  do not. This is confirmed by the measurement results with a signal to noise ratio of $\sim$ 4:1. From these measurements, we determine the expectation values of the observables: $\left\langle M_{k}^{\otimes 8}\right\rangle=\left\langle\left(\mbox{cos}\theta\sigma_x+\mbox{sin}\theta\sigma_y\right)\right\rangle$  which yields an average value of $0.610\pm0.026$  (see Fig. \ref{Fig3}b).

We can determine the fidelities of the cat states and detect the presence of genuine multipartite entanglement \cite{Bourennane04} using the tool of entanglement witness. The fidelity is a measure of to what extent the desired state is created and can be calculated by the overlap of the experimentally produced state with the ideal one. For the cat state (\ref{eq:GHZ}),
\bea
O_8 & = & \left|SC_8\right\rangle\left\langle SC_8\right| \nonumber\\
        & = & \frac{1}{2}\left(\left(\left|H\right\rangle\left\langle H\right|\right)^{\otimes 8}+\left(\left|V\right\rangle\left\langle V\right|\right)^{\otimes 8}\right)\nonumber\\
        &     & +\frac{1}{16}\sum_{n=0}^7(-1)^k(M_k)^{\otimes 8}
\label{eq:witness}
\eea
From the experimental data shown in Fig.\ref{Fig3}, we calculate the fidelity of our eight-photon Schr\"odinger-cat state: $0.708 \pm 0.016$. For the cat-type entangled states, it is sufficient to prove the presence of genuine multipartite entanglement if their fidelities exceed the threshold of 0.5. Thus, with a high statistical significance ($\sim$ 14 standard deviations), the genuine eight-photon entanglement is confirmed experimentally.

In conclusion, by exploiting the new techniques of ultra-bright entangled photon source, noise-reduction multi-photon interferometer and post-selection detection, we have experimentally generated and characterized the eight-photon Schr\"odinger-cat state. Being able to entangle eight individual single photons pushes forward the state-of-the-art six-photon  \cite{Lu07,Prevedel09,Wieczorek09,Radmark09,Matthews10} capability and will enable new quantum optics and quantum information processing experiments with multi-photon entanglement in previously inaccessible parameter regimes. One immediate application is to demonstrate the topological error correction scheme \cite{Raussendorf07} with eight-photon graph states. Furthermore, our eight-photon setup can serve a well-controlled few-qubit quantum simulation testbed for studying interesting phenomena in solid-state physics \cite{Ma11}, quantum chemistry \cite{Lanyon10}, and even biophysics \cite{Cai10}. Finally, it should also allow tests of the stability and dynamics of different families of entangled states (such as Schr\"odinger-cat states and one- and two-dimensional cluster states) under the effect of decoherence \cite{Barreiro10}, which may provide new insights into our understanding of the intriguing questions of classical to quantum transition.

We acknowledge M. Cramer for useful discussions. This work was supported by the National Natural Science Foundation of China, the Chinese Academy of Sciences and the National Fundamental Research Program (under Grant No: 2011CB921300). This work was also supported by the Alexander von Humboldt Foundation, the ERC.

\iffalse
\bibliographystyle{my4author1}
\bibliography{8Pho}
\fi

\clearpage

\section*{Appendix}
\renewcommand{\thefigure}{A\arabic{figure}}
 \setcounter{figure}{0}
\renewcommand{\theequation}{A.\arabic{equation}}
 \setcounter{equation}{0}
 \renewcommand{\thesection}{A.\Roman{section}}
\setcounter{section}{0}

\section*{Reducing noise contribution from double-pair emission}

\begin{figure}[htp]
\includegraphics[width=1.6\linewidth]{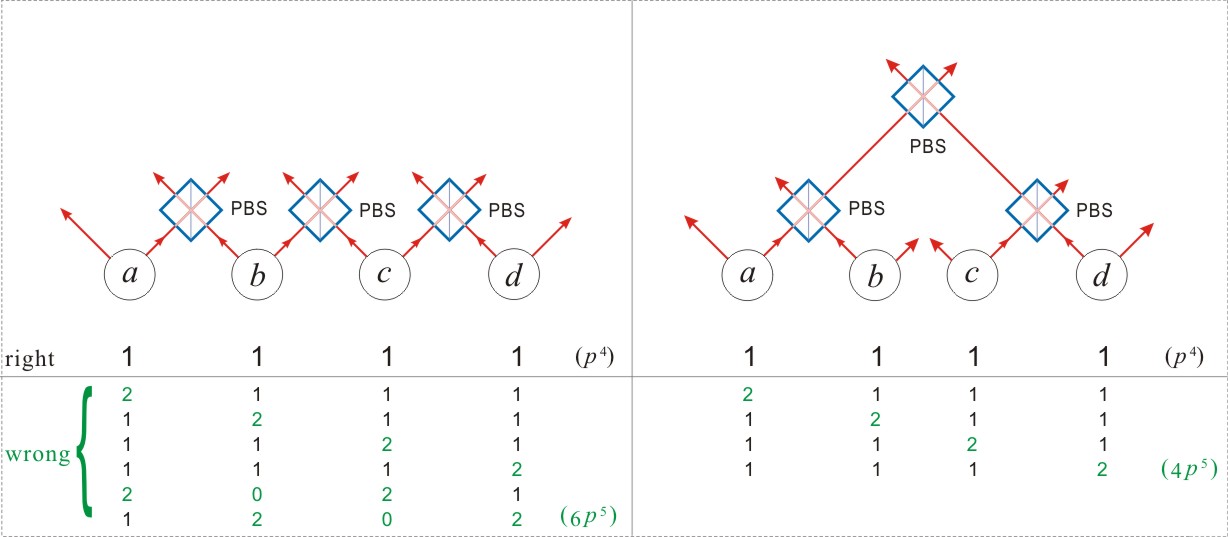}
\vspace{-0.1cm}\caption{Two possible configurations are shown for generating eight-photon entanglement. In the left panel the two-photon entangled pairs are connected successively in a chain topology (which was used in e.g. refs. \cite{Zhang06,Lu07a}), while in the right panel they are combined in a star topology. To generate the eight-photon Schr\"odinger-cat state, the desired case is that each of the inputs $a$, $b$, $c$ and $d$ has one and only one photon pair, the probability for which to happen is $p^4$  with p being the down-conversion probability. Also listed below the diagrams are the contributions from the double pair emission that will lead to erroneous eight-fold coincidence. We note that in the chain topology the total erroneous contribution is $6p^5$, while in the star topology it is $4p^5$, which suffers from less noise and is thus adopted in our present experiment. \label{FigS1}}
\end{figure}

\end{document}